\documentclass[11pt]{article}
\pdfoutput=1

\usepackage{amsfonts,amsmath,amssymb,graphicx}

\newcommand{\recov}{\Re}

\newcommand{\I}{\mathrm{i}}
\newcommand{\pr}{\mathbf{P}}
\newcommand{\ex}{\mathbf{E}}

\newcommand{\notthis}[1]{}

\newcommand{\half}{\frac{1}{2}}

\newcommand{\shalf}{{\textstyle\frac{1}{2}}}

\begin{document}

\title{\bf Emerging Market Corporate Bonds as First-to-Default Baskets}

\author{Richard Martin, Tolga Uzuner and Yao Ma}
\maketitle

\begin{abstract}
Emerging market hard-currency bonds are an asset class of growing importance, and contain exposure to an EM sovereign and the underlying industry. The authors investigate how to model this as a modification of the well-known first-to-default (FtD) basket, using the structural model, and find the approach feasible.

This is the longer version of an article published in RISK, June 2018.
\end{abstract}

Emerging market (EM) hard-currency corporate bonds have come to be regarded as a mainstream asset class over the last few years, and at the time of writing the market size is over 1~trillion USD \cite{IIF16}.
One reason for their popularity is that they provide exposure to corporate (sector) and country risk in a single package, i.e.~a double source of yield.
In the evaluation of investment opportunities in the asset class, there arises the fundamental question of how this risk combination should be done.

As an example, if a BBB$-$ Peruvian telecom firm comes to market with a new bond issue, at what yield or spread should it trade?---a question that needs to be answered if one is deciding whether or not to invest. In developed markets (DM) the sector risk can be obtained by fitting yield curves to all USD-denominated DM telecom bonds: with well over a hundred bonds (and many more than that for financial names), this is a relatively straightforward task. Similarly the risk of Peru can be understood through its USD-denominated sovereign bonds and/or CDS curve.
However, there may be no Peruvian telecom bonds of the required rating or maturity, so the bond market provides no direct answer.

Another application is in the evaluation of existing bonds. One component of the expected return for a particular credit is the expected movement in credit spread.
If a name is rated BB and we think its credit metrics are better than this, we can give it a rating of BB+ or better, so that we are predicting its spread to tighten; similarly if we dislike it then we can rate it BB$-$ or worse. To do this, we need spread curves for each rating, for a given (country,sector) pair.

As we said before, in some sense an EM corporate bond contains two risks embedded in one package, so it is tempting to add the spreads of the `standalone corporate' (DM sector) and the country, e.g.\ Peru~7y = 120bp, Telecom~7y~BBB$-$ = 145bp, so Peru telecom 7y BBB$-$ should trade at 265bp. 
However, this gives an answer that is almost always much too high, and is not workable.

Although the idea is completely non-standard in EM, one might try to think of the corporate as some sort of first-to-default (FtD) basket\footnote{See \cite{OKane08} for a general discussion on credit derivatives.}. The FtD spread is the sum of the constituent spreads if the underlying default risks are independent, but less if they are dependent, and so increasing the correlation might to do the trick. However, this is also destined to fail, even if less badly than the sum-of-spreads idea, for the simple reason that there are plenty of EM corporates that trade comparably with or tighter than their sovereign, whereas an FtD spread must be at least that of the widest constituent.
As examples:
\begin{itemize}
\item
Brazil: Brazil Foods (BRFSBZ) until recent concerns, Braskem, Embraer, Globo Communica\c{c}\~ao (GLOPAR), Klabin, Vale;
\item
Mexico: Mexichem (MXCHF);
\item
Russia: Severstal (CHMFRU), Norilsk (GMKNRM), Lukoil;
\item
Turkey: Coca-Cola Icecek (CCOLAT), Tupras (recently). 
\end{itemize}
Most of these examples have higher rating than the sovereign, except Mexichem.
Typically these have businesses that are strong enough to withstand a sovereign default, so it is incorrect to argue that a sovereign default must trigger a corporate default.
Accordingly, the FtD construction necessarily overestimates the corporate's spread in these cases, and in others too.

Nonetheless, we suggest that there is some mileage in the idea that an EM corporate is a sort of FtD, but it requires judicious reformulation, and the model that we introduce next describes this in detail.

\section{Model}

\subsection{Credit barrier model}

We establish a credit barrier model in which default is caused by the passage of a process $X_t$ (notionally the logarithm of the discounted firm value $e^{-rt} A_t$, appropriately normalised so that it starts from zero) below a barrier. The dynamics of $X_t$ are a diffusion of volatility $\sigma$ added to  exponentially-distributed down-jumps of rate $\lambda$ and mean $-\xi$. The first expression below ($X^a_t$) is for the standalone corporate, and the second for the country ($X^c_t$):
\begin{eqnarray}
dX^a_t &=& \sigma_a \, dW^a_t - \xi_a \, dJ^a_t + \mu_a\, dt \label{eq:Xat} \\
&& \mbox{\!\!\!\!\!\!\!\!\!\!\!\! Correl $\rho$} \updownarrow \;\;\;\;\;\;\;\;\;\;\;\; \updownarrow\mbox{Independent} \nonumber \\
dX^c_t &=& \sigma_c \, dW^c_t - \xi_c \, dJ^c_t + \mu_c\, dt \label{eq:Xct} 
\end{eqnarray}
with $W_t^{a,c}$ standard Brownian motions.
The rationale for this is that the Brownian terms capture global market moves---and, as spreads are highly correlated, so too must be the firm values, necessitating a high correlation $\rho$---whereas the jumps represent idiosyncratic effects.
The firm value $A_t$ is, at least in principle, a tradable asset, so its discounted expectation, which is $e^{X_t}$, should be a martingale under the pricing measure $\mathbf Q$. This necessitates
\[
\mu_a = -\shalf \sigma_a^2 + \frac{\lambda_a \xi_a}{1+\xi_a}, 
\qquad
\mu_c = -\shalf \sigma_c^2 + \frac{\lambda_c \xi_c}{1+\xi_c}. 
\]

The barrier is a constant ($-L$ say) for each\footnote{The idea is that the debt burden grows at the riskfree rate $r$, in the same way that the firm value does (under $\mathbf Q$).}, and in view of the fact that we can rescale $\sigma,\xi,L$ without affecting the result\footnote{This is not quite correct, because to obtain identical results on rescaling we would need $\mu$ to scale linearly with $\sigma,\xi$; however, the martingale condition contradicts this. Nonetheless this is a minor consideration in practice. In the context of a firm value model, we set it up so that $L=\ln(A/B)$ where $A$ is the firm value, and $B$, the value of the firm at which default occurs, is related to the indebtedness or leverage of the entity. Upon calibration to the term structure of default, $\sigma,\xi$ tell us the riskiness of the firm as imputed from the credit market. In effect by setting $L=1$ we are circumventing the business of estimating the entity's leverage.} we set $L=1$, at least in the first instance.
The default times are therefore given by
\begin{eqnarray}
\tau_a &=& \inf_T \Big( \min_{0\le t \le T} X^a_t < -1 \Big) \label{eq:taua}
\\
\tau_c &=& \inf_T \Big( \min_{0\le t \le T} X^c_t < -1 \Big)
\label{eq:tauc}
\end{eqnarray}
From this the par CDS spread can be obtained from the cumulative default probability $P(t)$ in the usual way\footnote{Approximating the coupon leg as continuous.}:
\[
s(T) = \frac{(1-\recov) \int_0^T B^\circ(t) \, dP(t)}{\int_0^T B^\circ(t) \big(1-P(t)\big) \, dt}
\]
where $\recov$ denotes the recovery (set at $40\%$ throughout) and $B^\circ$ is the riskfree discount factor. The numerator and denominator are respectively the PV of the default leg and the risky PV01.

It is worth making some remarks on the way the model is set up and in particular the number of parameters. Credit barrier models have, at least in principle, infinitely many degrees of freedom as regards asset dynamics (term structure of volatility) and barrier shape. It is clear that, unless one distorts the term structure of volatility or the barrier in an unrealistic way, down-jumps are necessary to give a realistic shape of term structure, so this adds further complexity.
While there are situations in which such flexibility is desirable, in `normal market conditions' parameters should not be time-varying, but even then a basic difficulty is that there are too many parameters. If in the above we allow $\sigma,\lambda,\xi$ to be variable, we have a degenerate model because when $\lambda$ is large and $\xi$ is small we can trade the diffusion term off against the jump term, as there is no difference between many tiny jumps and a diffusion process. This point was discussed \cite{Martin09a,Martin10b} in the context of the Carr-Madan-Yor model, wherein it is argued that the right number of parameters is two. One way around this is to fix $\lambda$ according to the following rating-dependent scheme:
\begin{equation}
\begin{tabular}{lr}
\hline
& $\lambda$ (per yr) \\\hline
A & 0.125 \\
BBB & 0.25 \\
BB & 0.5 \\
B & 1 \\
\hline
\end{tabular}
\label{eq:lambdas}
\end{equation}
The jump term is thereby restricted to providing occasional and potentially large negative jumps---for example a BBB rated firm suffers a jump every four years on average---and the diffusion term captures general market movements. As, across different issuers, the diffusions are highly correlated, one effect of this is to make the spreads of high-rated firms highly correlated, which is in accordance with experience: Latam banks are one of the best examples in recent years. High-yield firms are much more susceptible to idiosyncratic problems, and so have a higher density of idiosyncratic jumps. 
The parameters $\sigma,\xi$ do different things because jumps affect the short end of the curve whereas the diffusive component does not.

The probability distribution of the first passage time $\tau$ can in principle be evaluated by the Wiener-Hopf factorisation, following standard arguments on spectrally-negative L\'evy processes (see e.g.~\cite{Sato02, Madan08}). However, the results are not closed-form for this model, requiring an inverse Laplace transform, and in any case what we are about to do with the FtD construction cannot be done analytically. For the purposes of this paper we have fallen back on Monte Carlo simulation, using 100,000 simulations in all cases. This is sufficient to give a standard deviation of $\lesssim0.5\%$ of spread, which is more than enough precision for the current work. The number of simulations can be reduced by using importance sampling (for an introduction see e.g.~\cite{Glasserman04}).

\subsection{Calibration}

We have already said that in calibration to a single credit, there are two free parameters $(\sigma,\xi)$ and that these are easily found. This allows the country parameters $(\sigma_c,\xi_c)$ to be found.

For the standalone corporate, what one typically needs is a parameter set per rating, and CDS curves are not always available, so calibration to bonds may be needed. An important complication is that corporates do not necessarily trade in line with their rating---rather they trade in line with where the market thinks they should be rated---and so we need to take a large number of bonds to iron out these idiosyncratic effects. As broad sectors in DM contain well over a hundred bonds (and banks/financials many more than that) this is workable. However, the analytical intractability of the model makes it difficult to calibrate.

We use the following two-step method.
First, fit a simple `model' (really, just a parameterisation) of the available DM corporate bond or CDS spread data, indexed by rating $k$:
\[
s_k(T) = e^{a_k} + e^{b_k-\theta T}.
\]
We use one parameter pair $(a_k,b_k)$ for each `broad rating' (A, BBB, BB, B) and for a bond that has a rating modifier (e.g.\ BBB+) we linearly interpolate between the adjacent broad ratings. Thus, in fitting a given sector, we have nine parameters: four $(a_k,b_k)$ pairs, and $\theta$. Each parameter does something materially different from the others, in that $e^{a_k}+e^{b_k}$ is the short-term spread and $e^{a_k}$ is the long-term spread for rating $k$, so the fitting is unambiguous. It is also very fast, taking a second or so per sector when using the JP~Morgan JULI and JPDO indices, which respectively cover the investment grade and high yield DM universe (ca.~8000 bonds in all).

The next step is to take each fitted curve in turn, including those for modified ratings A$-$, BBB+, etc., and find the credit barrier model parameters $\sigma_a,\xi_a$ that correspond most closely to it. The objective function is the worst of the fitting errors for the three tenors 2y, 5y, 10y.
We have, in fact, been able to tighten the parameterisation by using the same value of $\xi$ (but different $\sigma$'s) for each corporate rating, depending on the sector and date: so in Table~\ref{tab:params} later on, where we present real examples, in each row the $\xi_c$'s are the same.

\notthis{
There is, incidentally, a reason for fixing $\lambda_a$ and allowing $\xi_a$ (and $\sigma_a$) to move during this fitting process, rather than fixing $\xi_a$ and varying $\lambda_a$. The former allows recalculation of the cumulative default probability without having to resimulate, because we can simulate paths of a unit Brownian motion $W_t$ and of a jump process $J_t$ with jumps of unit mean, and then combine them in appropriate proportions, dictated by $\sigma_a,\xi_a$, to obtain paths of $X^a_t$. The latter requires resimulation of the jump process, with different jump rate, each time $\lambda_a$ is varied.
} 

\notthis{
The estimated parameters are obtained by minimising an objective function quantifying the closeness of fit. At each evaluation of this, we do not want to resimulate the processes $W^{a,c}_t$ and $J^{a,c}_t$. However, we do not have to, because we can write
\[
\left. \begin{array}{l} W^a_t \\ W^c_t \end{array} \right\}
 = \sqrt{\frac{1+\rho}{2}} \, W^+_t \pm \sqrt{\frac{1-\rho}{2}} \, W^-_t
\]
where $W^\pm_t$ are independent Brownian motions,
and then simulate a set of realisations of $W^{\pm}_t$ and $J^{a,c}_t$. Then the necessary processes $X^{a,c}_t$ are obtained by linearly combining these using $\rho$, $\sigma$, $\xi$. Upon changing any of these parameters no resimulation need be done: it is just a question of recombining the simulations.
This, incidentally, is the reason for fixing $\lambda$ rather than $\xi$ at the outset: were $\lambda$ to be changed, resimulation of $J^{a,c}_t$ would be needed.
} 

\subsection{FtD and Modified FtD}

In a simple FtD, the default of the EM corporate bond ($b$) in question would just be the minimum of the two default times:
\[
\tau_b = \tau_a \wedge \tau_c.
\]
Incidentally as we the short-term is determined entirely by jumps, and as (\ref{eq:Xat},\ref{eq:Xct}) makes these independent, the short-term FtD spread  is in this model just the sum of the spreads.

However, as we said above, this causes the EM corporate to trade at least as wide as its country; put differently, the country has too important an effect. To fix this problem we keep (\ref{eq:taua},\ref{eq:tauc}), but redefine $\tau_b$:
\begin{equation}
\tau_b = \tau_a \wedge \inf_T \Big( \min_{0\le t \le T} X^c_t < - L^*_c \Big)
\label{eq:taub}
\end{equation}
where now $L^*_c \ge 1$. The greater is $L^*_c$, the worse the country has to get to cause a default of the corporate, and the less is the contribution of the country spread to the corporate spread. In the limit $L^*_c\to\infty$ the country has no effect at all and we are back to standalone risk: $\tau_b=\tau_a$.
We call this the `modified FtD' (MFtD).

We argue that the value of $L^*_c$ should depend on the corporate's credit quality. A very highly-rated credit (e.g.~AA) is expected to perform well almost regardless of its country because its business model, asset base and funding sources are assumed to be well-diversified. Conversely a poor-quality credit (e.g.~CCC) is much more susceptible to problems with the country it is in, additionally to idiosyncratic and sectoral problems: such names typically have negative free cash flow and/or low EBITDA.
The scheme shown below achieves this:
\begin{equation}
\begin{tabular}{lr}
\hline
Rating & $L^*_c$ \\
\hline
A & $1.45$ \\
BBB & $1.35$ \\
BB & $1.20$ \\
B & $1.00$ \\
\hline
\end{tabular}
\label{eq:Lc}
\end{equation}
and in the examples below we have used this same set of values throughout.

The parameter $\rho$ has no effect on the single-name calibrations but it does affect the MFtD price. We have fixed it at 80\%, as this is the approximate correlation between corporate sectors and countries, estimated historically\footnote{By correlating monthly log-returns of each CEMBI (corporate) subsector with the EMBI (sovereign index), since 2004. Data provider: JP~Morgan.}.

\subsection{Summary of calibration procedure}

To summarise:
\begin{itemize}
\item
$\sigma_c,\xi_c$ are determined from the country CDS/cash curve;
\item
$\sigma_a,\xi_a$ are determined from developed market cash bonds;
\item
$\lambda_{a,c}$ are fixed using the rating-dependent values in (\ref{eq:lambdas});
\item
$\mu_{a,c}$ are determined by no-arbitrage conditions;
\item
$\rho$ is fixed at 80\% throughout;
\item
the only determinant of EM corporate spreads, once these parameters have been fixed from the country curve and the DM corporate curves, is $L^*_c$ given in (\ref{eq:Lc}).
\end{itemize}

\section{Examples and Discussion}

\subsection{Comparison with market data}

The chosen sectors/countries in our study are:
\begin{center}
\begin{tabular}{lll}
\hline
Figure & Sector & Country \\
\hline
(1,2) & Food & Brazil \\
(3,4) & Mining & Peru \\
(5,6) & Oil\&Gas & Russia \\
(7,8) & Banks & Turkey \\
(9,10) & Real Estate & China \\
\hline
\end{tabular}
\end{center}

The country curve is the black dotted line, the dashed lines show the standalone curves, colour coded by credit rating, and the solid lines show the modelled EM corporate curves using the same colour scheme. To avoid cluttering up the plot we have only plotted the main ratings A, BBB, BB, B, omitting the fine grades (A--, BBB+ etc.).
Corporate bond spreads are marked on the plots, and are to be compared with the solid curves of corresponding rating.

Each is done on two different dates: Figures~1,3,5,7,9 are in January 2016 and Figures~2,4,6,8,10 in October 2017 when spreads were much tighter across the board. Some of the bonds were upgraded over the intervening period, and a smaller number were downgraded.
The calibration parameters are shown in Table~\ref{tab:params} and the results are graphed in the corresponding figure number.
In some cases the bonds are callable so the spread-to-worst is being used. For example, HOCLN~'21 is trading to-maturity in Figure~3, but to its Jan'18 call date in Figure~4; similarly with LNGFOR~'23 in Figures~9,10.

Agreement is reasonably good: thus in Figure~1, BRFSBZ lies slightly outside the BBB curve, while BEEFBZ lies in between the BB and B curves, as one would hope.



The agreement for Russian oil/gas names is not bad, but the model overestimates the spread a little (Figure~5). This is mainly because the ratings that are applied to them are rather punitive and caused by the sovereign: many of these names would be rated a few notches higher if they were in developed markets. The right way to think about those rated BBB$-$/BB+ is that they should be regarded as, roughly, BBB$+$/BBB standalone oil\&gas risk combined with Russia risk, not BBB$-$/BB+ standalone oil\&gas risk combined with Russia risk.

Another source of disagreement is Turkish banks, though in Figure~8 rather than Figure~7. Unlike the other examples, it is clear that Turkish banks have not followed the market tighter to anywhere near the same extent than other sectors between Jan'16 and Oct'17. Part of this is due to the attempted coup in Turkey in late 2016. The market has repriced the risk and, despite no change in the credit ratings, it regards them as trading a couple of notches wider (i.e.~BB$-$ rather than BB$+$).
Another example is high-yield Chinese real estate, where the modelled BB and B curves in Figures~9,10 are too tight relative to the bonds, though the A and BBB curves are fine.
This brings us conveniently to a couple of extensions to the model, which we discuss now.

\subsection{Extensions 1: Risk priced differently in EM}

We are assuming that the purely corporate part of the EM risk is priced the same in emerging and developed markets. This may not always be so, for a couple of reasons. First, a general selloff in EM, caused perhaps by a reallocation of capital away from EM assets back into DM, will cause EM assets to price wider.
Secondly, it may be something specific to the sector, notable examples of which are Turkish banks and high-yield Chinese real estate, as just discussed. If the market regards a particular EM sector as fundamentally riskier or more opaque than its DM counterpart, then it will naturally price at a higher spread.
Either way there is something more to the pricing of EM risk than a simple combination of DM corporate risk and EM sovereign risk.

In the structural model the most convenient way to increase the default probability, and hence the credit spread, is to increase the volatility---or we could move the default barrier closer, which is in effect the same.
In view of this, the suggested solution is to redefine $\tau_b$ as
\begin{equation}
\tau_b = \inf_T \Big( \min_{0\le t \le T} X^a_t < - L^*_a \Big) \wedge \inf_T \Big( \min_{0\le t \le T} X^c_t < - L^*_c \Big)
\label{eq:taub2}
\end{equation}
where now $L^*_a$, previously unity, is made $\le1$ so as to push the default probability higher.
In the Turkish banks example, adjusting $L^*_a$ from $1$ to $0.85$ has the desired effect: see Figure~8(iii). The same alteration, applied only to the BB and B curves of Chinese real estate, has the same beneficial effect: see Figure~11.

\notthis{
We have used the following modifications:
\begin{center}
\begin{tabular}{lr}
\hline
Rating & $L^*_a$ \\
\hline
A & $1.00$ \\
BBB & $0.95$ \\
BB & $0.90$ \\
B & $0.85$ \\
\hline
\end{tabular}
\end{center}
and plotted the results in Figure 3(a,b) for Chinese real estate. The agreement is generally better. In Figure~3(b), it is clear that the Evergrande (EVERRE) curve is steeper than that of a typical B$-$ credit, particularly for longer maturities.
In fact this is for reasons that are idiosyncratic to the issuer: the amount of debt outstanding is small for the short-maturity bonds, but much higher at the long end\footnote{Up to 5y maturity the bond issues are a few hundred \$M each, but the 2025s are almost \$5000M.}, so the longer-dated debt is likely to trade wider.
} 


\subsection{Extensions 2: Quasi-sovereigns}

Another possible modification is for quasi-sovereigns: that is to say, names that are largely state-owned and therefore have strong correlation with their sovereign.
There is a free option on the part of the sovereign issuer to issue debt out of the quasi-sovereign entity without properly guaranteeing it, and so the quasi-sovereign should trade wider than the sovereign. In principle it may trade tighter than the DM sector of the same rating, as for example Codelco (CDEL, Chile) did in 2016.
This suggests that the opposite construction to (\ref{eq:taub}) is needed, in the sense of:
\begin{equation}
\tau_b = \inf_T \Big( \min_{0\le t \le T} X^a_t < - L^*_a \Big) \wedge \tau_c
\label{eq:taub3}
\end{equation}
so that the name has full exposure to the country, but only partial exposure to the industry if $L^*_a>1$.

\section{Conclusion}

We have presented an arbitrage-free model for understanding the compound risk present in EM corporate bonds and demonstrated its use in several market sectors with recent data, where it works tolerably well.

We could in principle have formulated the model differently, by directly modelling the default time probabilities. However, these kind of models seem to us to look more like mathematical trickery rather than an attempt to model what is really going on.
In a FtD, one has
\[
\pr(\tau_b < t) = \pr (\tau_a < t \cup \tau_c < t)
\]
and for the MFtD one could reduce the impact of the country by writing instead
\[
\pr(\tau_b < t) = \pr (\tau_a < t \cup \tau_c < \eta t), \qquad \eta<1
\]
though why this should be the right construction is unclear.
The bivariate probability is then obtained using the univariate probabilities (obtained in the usual way from the spread curves of the country and DM sector) and the now-infamous copulas. Certainly such a model would be easier to calculate, but it loses any connection with the fundamental origins of credit, which the credit barrier models capture quite well. And, as always, there is the question of what copula to use.

The comment we have just made about computation brings us back to the point we made earlier about calculating the hitting probabilities in the credit barrier model. As we said earlier, the univariate ones can be calculated by Wiener-Hopf, by consideration of the generating function
\[
F(u,s) = \int_0^\infty \ex[e^{\I u X_t}] e^{-st} \, dt,
\]
which can be evaluated easily enough for most well-known L\'evy processes; factorisation of $s/(s-F)$ gives information about the first passage time to a boundary, after inversion of a Laplace transform \cite{Sato02}. For a bivariate problem, i.e.\ understanding the first passage time of the pair $(X^c_t,X^a_t)$, one can calculate the two-dimensional generating function
\[
F(u,v,s) = \int_0^\infty \ex[e^{\I u X^a_t + \I v X^c_t}] e^{-st} \, dt,
\]
for the model (\ref{eq:Xat},\ref{eq:Xct}), but a considerable amount of work is needed to extract the information about the stopping time $\tau_a \wedge \tau_c$.
In general the computation of first passage times for all but the simplest models is still a difficult challenge in, and outside, quantitative finance.

We end by making a wry observation that will amuse readers missing the structured credit boom of the mid 2000s: despite structured credit instruments apparently having fallen into disrepute and disuse since 2008, a simple form of FtD (or, at least, a modified form of it) has in fact been alive and well, and trading in the form of EM corporate bonds!

\vspace{5mm}
\emph{The authors are grateful to Jackie Choi, Stuart Firth, Belinda Hill, John Lam, Nuno Pinto,  Michael Story and Rodney Thomas for helpful discussions.}

\bibliographystyle{plain}
\bibliography{../phd}

\begin{thebibliography}{1}

\bibitem{Glasserman04}
P.~Glasserman.
\newblock {\em Monte Carlo Methods in Financial Engineering}.
\newblock Springer, 2004.

\bibitem{Madan08}
D.~P. Madan and W.~Schoutens.
\newblock Break on through to the single side.
\newblock {\em J. Risk}, 4(3):3--20, 2008.

\bibitem{Martin09a}
R.~J. Martin.
\newblock Credit spread shocks: {H}ow big, and how often?
\newblock {\em RISK}, 22(8):84--89, 2009.

\bibitem{Martin10b}
R.~J. Martin.
\newblock Smiling {J}umps.
\newblock {\em RISK}, 23(9):108--113, 2010.

\bibitem{OKane08}
D.~O'Kane.
\newblock {\em Valuing single-name and multi-name credit derivatives}.
\newblock Wiley, 2008.

\bibitem{Sato02}
K.-I. Sato.
\newblock {\em L\'evy Processes and Infinitely Divisible Distributions}.
\newblock Cambridge University Press, 2002.

\bibitem{IIF16}
E.~Tiftik, K.~Mahmood, H.~Tran, and S.~Gibbs.
\newblock {EM} {D}ebt {M}onitor.
\newblock Technical report, Institute of International Finance, 2016.
\newblock {\tt www.iif.com}.

\end{thebibliography}

\vspace{20mm}

\begin{table}[htbp]
\centering
\begin{tabular}{r|rr|rr|rr|rr|rr}
&&&
\multicolumn{2}{|c|}{A} & \multicolumn{2}{|c|}{BBB} & \multicolumn{2}{|c|}{BB} & \multicolumn{2}{|c}{B} \\
Fig. & $\sigma_c$ & $\xi_c$ & $\sigma_a$ & $\xi_a$ & $\sigma_a$ & $\xi_a$ & $\sigma_a$ & $\xi_a$ & $\sigma_a$ & $\xi_a$ \\
\hline
1 & 0.32 & 0.25 & 0.18 & 0.27 & 0.16 & 0.27 & 0.16 & 0.27 & 0.15 & 0.27 \\
2 & 0.25 & 0.13 & 0.16 & 0.25 & 0.14 & 0.25 & 0.11 & 0.25 & 0.14 & 0.25 \\
3 & 0.22 & 0.25 & 0.27 & 0.27 & 0.29 & 0.27 & 0.38 & 0.27 & 0.62 & 0.27 \\
4 & 0.20 & 0.17 & 0.16 & 0.24 & 0.14 & 0.24 & 0.14 & 0.24 & 0.17 & 0.24 \\
5 & 0.26 & 0.23 & 0.23 & 0.30 & 0.26 & 0.30 & 0.33 & 0.30 &   &   \\
6 & 0.18 & 0.17 & 0.16 & 0.25 & 0.14 & 0.25 & 0.15 & 0.25 &   &   \\
7 & 0.23 & 0.25 & 0.17 & 0.27 & 0.16 & 0.27 & 0.21 & 0.27 &  &   \\
8 & 0.25 & 0.14 & 0.15 & 0.23 & 0.15 & 0.23 & 0.15 & 0.23 &   &   \\
9 & 0.21 & 0.24 & 0.15 & 0.22 & 0.23 & 0.22 & 0.23 & 0.22 & 0.27 & 0.22    \\
10 & 0.17 & 0.20 & 0.18 & 0.21 & 0.18 & 0.21 & 0.18 & 0.21 & 0.21 & 0.21    \\

\end{tabular}
\caption{Calibration parameters. Brazil, Russia and Turkey all rated BB ($\lambda=0.5$/yr), Peru A/BBB+ ($\lambda=0.2$/yr), China A ($\lambda=0.125$/yr).
For corporates, the $\lambda$'s are given in eq.~(\ref{eq:lambdas}).}
\label{tab:params}
\end{table}

\begin{figure}[!h]
\centering
\begin{tabular}{ll}
(i) & \includegraphics[scale=0.9]{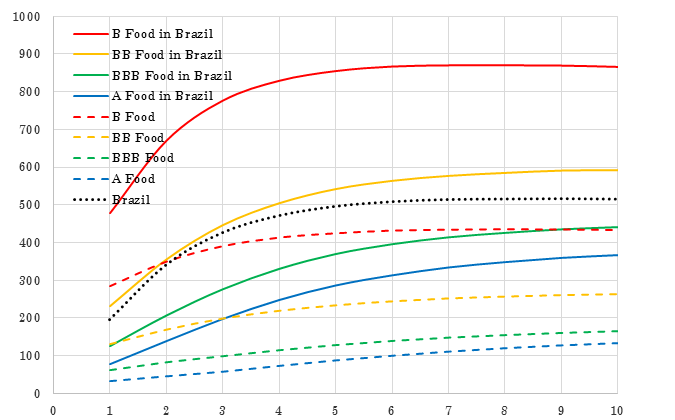} \\
(ii) & \includegraphics[scale=0.9]{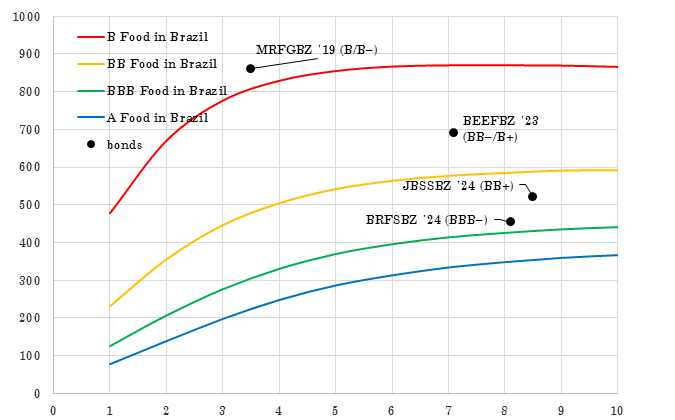} \\
\end{tabular}
\caption{Food/Brazil, Jan'16: (i) MFtD construction with DM curves, sovereign curve, and EM curves; (ii) EM curves with relevant bonds shown.}
\label{fig:1}
\end{figure}

\begin{figure}[!h]
\centering
\begin{tabular}{ll}
(i) & \includegraphics[scale=0.9]{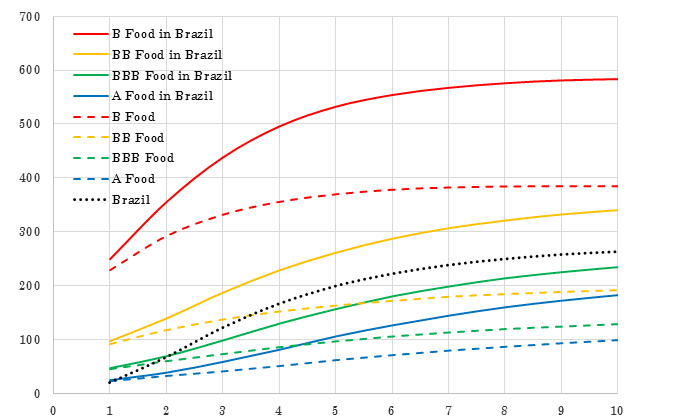} \\
(ii) & \includegraphics[scale=0.9]{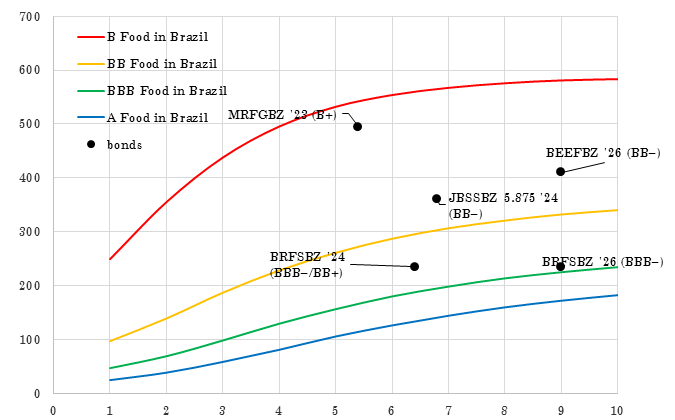} \\
\end{tabular}
\caption{Food/Brazil, Oct'17: (i) MFtD construction with DM curves, sovereign curve, and EM curves; (ii) EM curves with relevant bonds shown.}
\label{fig:2}
\end{figure}

\begin{figure}[!h]
\centering
\begin{tabular}{ll}
(i) & \includegraphics[scale=0.9]{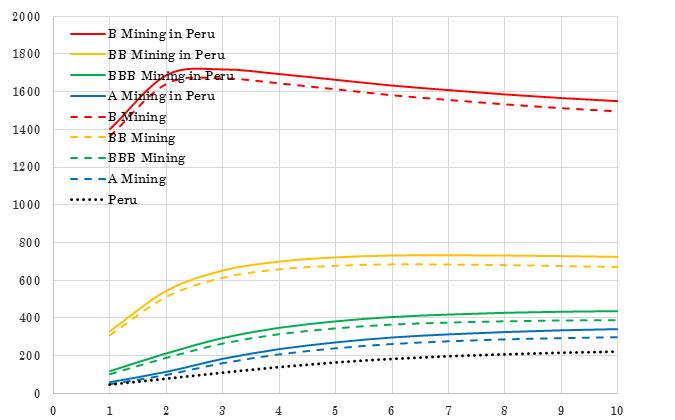} \\
(ii) & \includegraphics[scale=0.9]{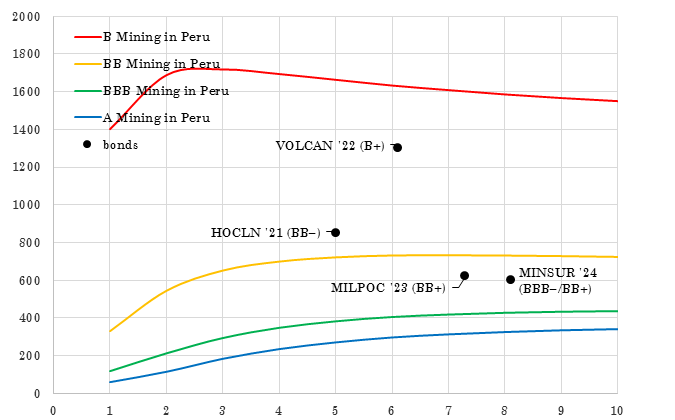} \\
\end{tabular}
\caption{Mining/Peru, Jan'16: (i) MFtD construction with DM curves, sovereign curve, and EM curves; (ii) EM curves with relevant bonds shown.}
\label{fig:3}
\end{figure}

\begin{figure}[!h]
\centering
\begin{tabular}{ll}
(i) & \includegraphics[scale=0.9]{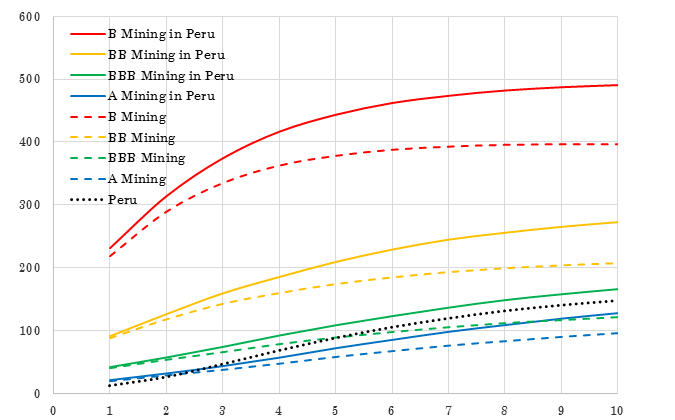} \\
(ii) & \includegraphics[scale=0.9]{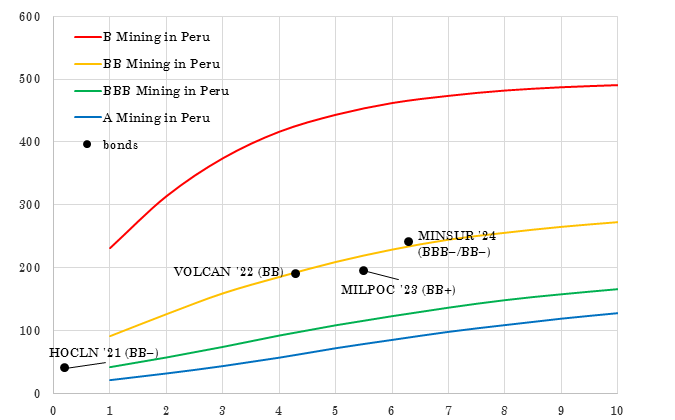} \\
\end{tabular}
\caption{Mining/Peru, Oct'17: (i) MFtD construction with DM curves, sovereign curve, and EM curves; (ii) EM curves with relevant bonds shown.}
\label{fig:4}
\end{figure}

\begin{figure}[!h]
\centering
\begin{tabular}{ll}
(i) & \includegraphics[scale=0.9]{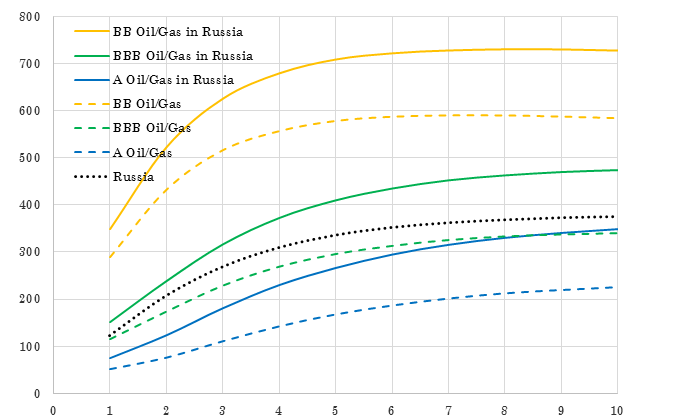} \\
(ii) & \includegraphics[scale=0.9]{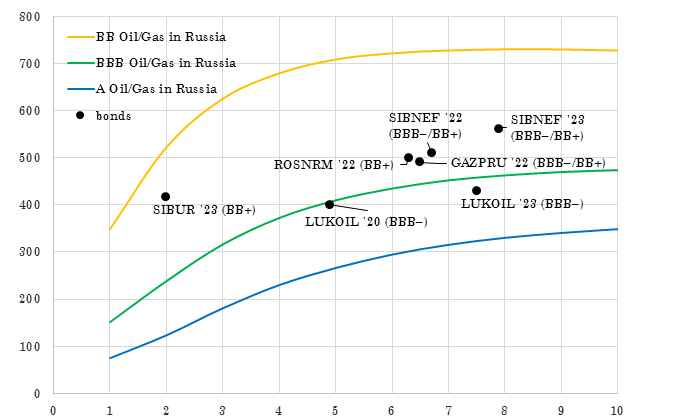} \\
\end{tabular}
\caption{Oil\&Gas/Russia, Jan'16: (i) MFtD construction with DM curves, sovereign curve, and EM curves; (ii) EM curves with relevant bonds shown.}
\label{fig:5}
\end{figure}

\begin{figure}[!h]
\centering
\begin{tabular}{ll}
(i) & \includegraphics[scale=0.9]{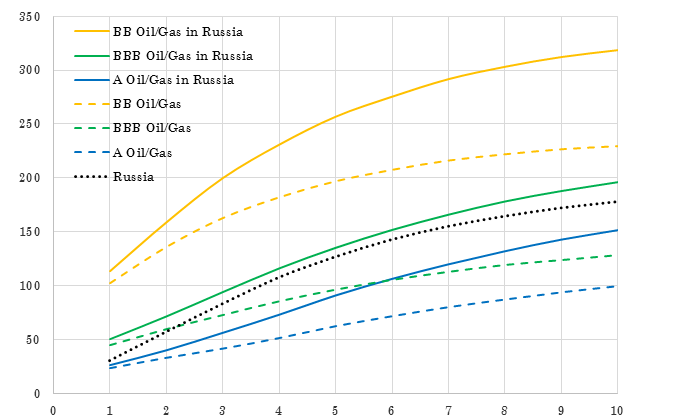} \\
(ii) & \includegraphics[scale=0.9]{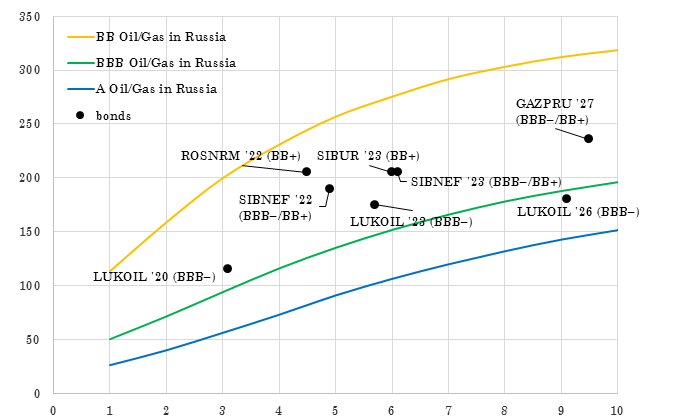} \\
\end{tabular}
\caption{Oil\&Gas/Russia, Oct'17: (i) MFtD construction with DM curves, sovereign curve, and EM curves; (ii) EM curves with relevant bonds shown.}
\label{fig:6}
\end{figure}

\begin{figure}[!h]
\centering
\begin{tabular}{ll}
(i) & \includegraphics[scale=0.9]{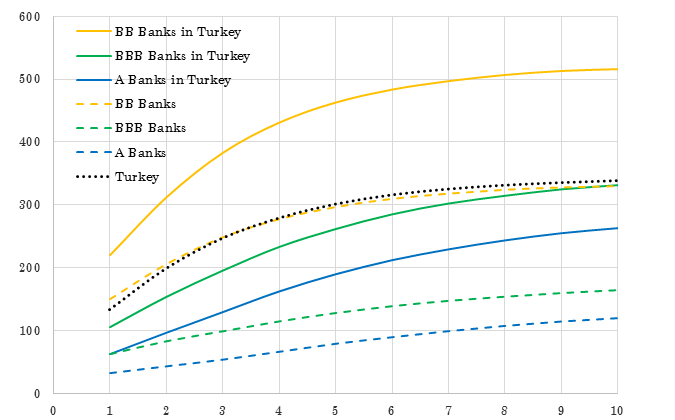} \\
(ii) & \includegraphics[scale=0.9]{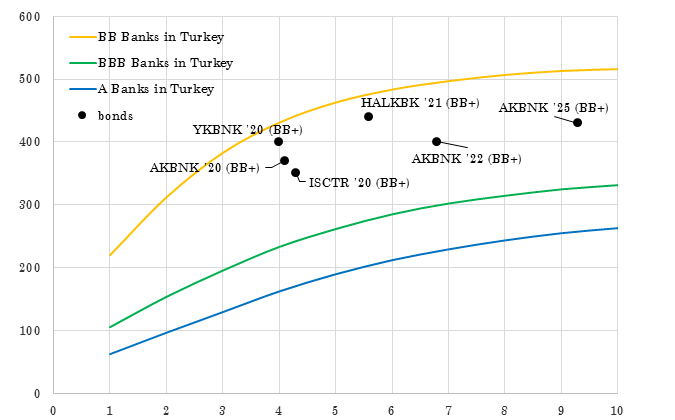} \\
\end{tabular}
\caption{Banks/Turkey, Jan'16: (i) MFtD construction with DM curves, sovereign curve, and EM curves; (ii) EM curves with relevant bonds shown.}
\label{fig:7}
\end{figure}

\begin{figure}[!h]
\centering
\begin{tabular}{ll}
(i) & \includegraphics[scale=0.9]{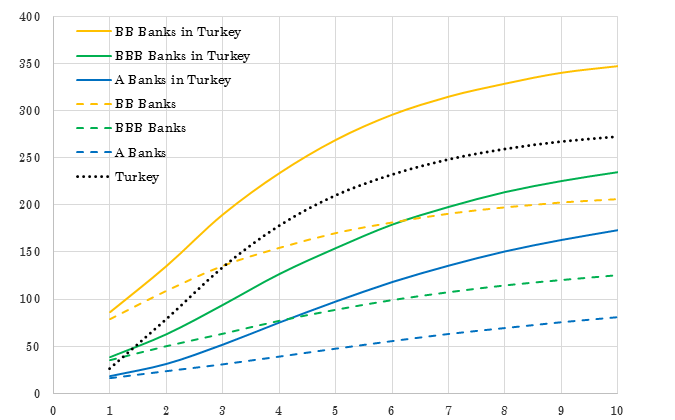} \\
(ii) & \includegraphics[scale=0.9]{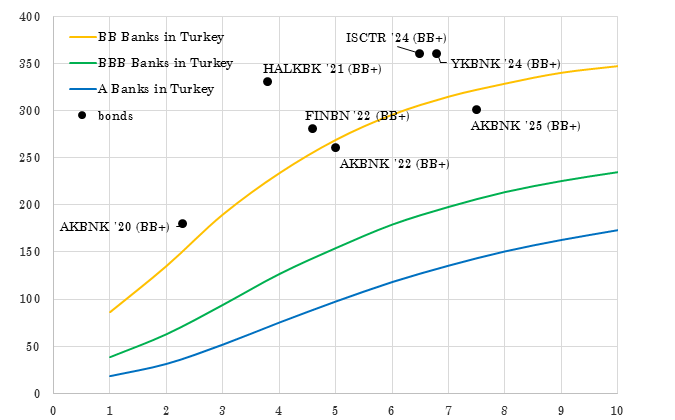} \\
\end{tabular}
\caption{Banks/Turkey, Oct'17: (i) MFtD construction with DM curves, sovereign curve, and EM curves; (ii) EM curves with relevant bonds shown.}
\label{fig:8}
\end{figure}

\addtocounter{figure}{-1}

\begin{figure}[!h]
\centering
\begin{tabular}{ll}
(iii)  & \includegraphics[scale=0.9]{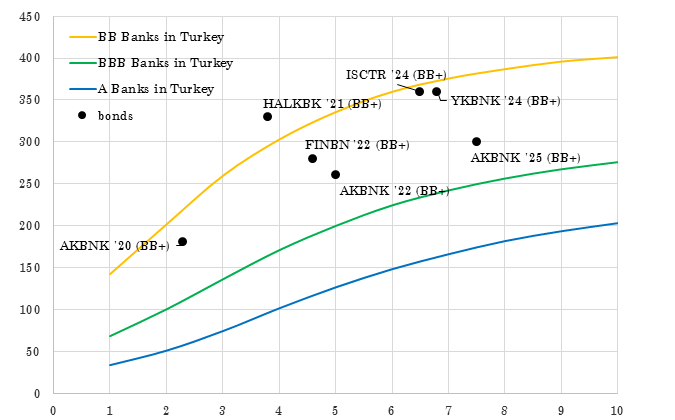} \\
\end{tabular}
\caption{Banks/Turkey, Oct'17: modified EM curves ($L_a^*=0.85$, see text) with relevant bonds shown.}
\label{fig:8a}
\end{figure}

\begin{figure}[!h]
\centering
\begin{tabular}{ll}
(i) & \includegraphics[scale=0.9]{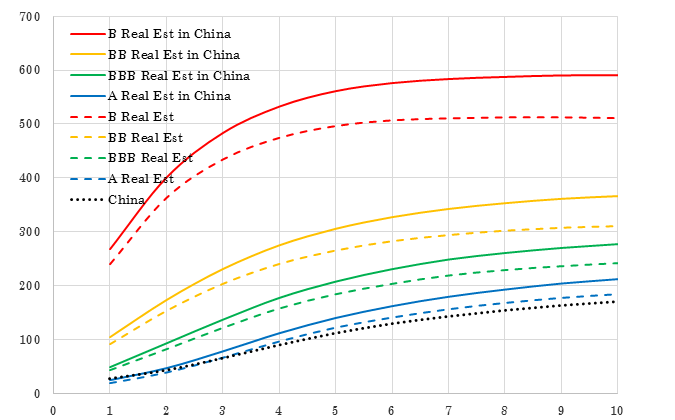} \\
(ii) & \includegraphics[scale=0.9]{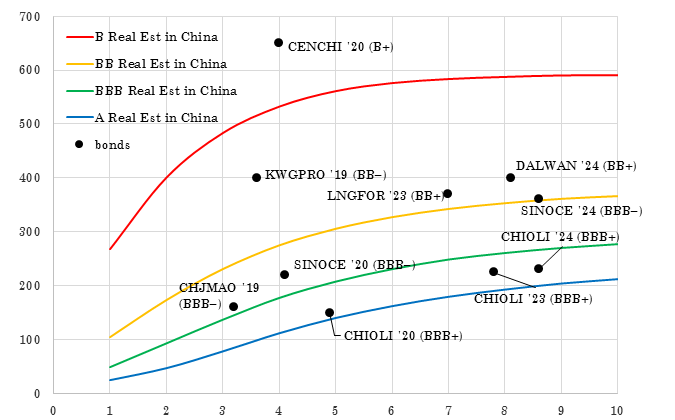} \\
\end{tabular}
\caption{Real Estate/China, Jan'16: (i) MFtD construction with DM curves, sovereign curve, and EM curves; (ii) EM curves with relevant bonds shown.}
\label{fig:9}
\end{figure}

\begin{figure}[!h]
\centering
\begin{tabular}{ll}
(i) & \includegraphics[scale=0.9]{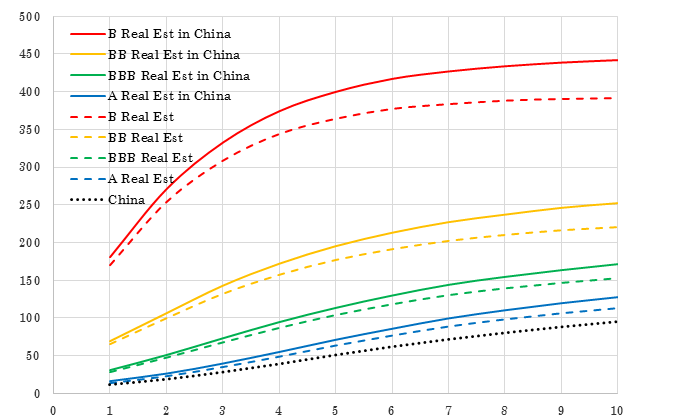} \\
(ii) & \includegraphics[scale=0.9]{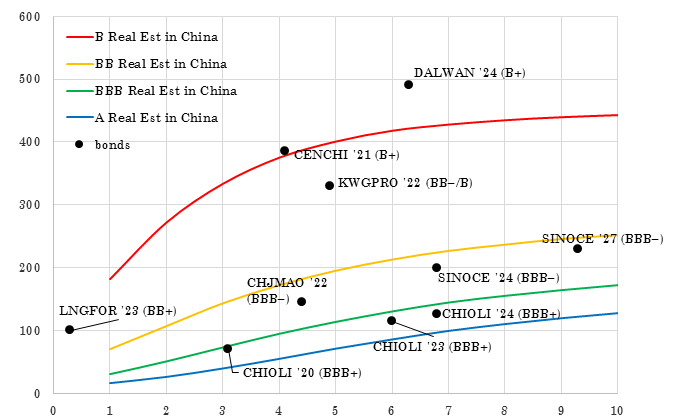} \\
\end{tabular}
\caption{Real Estate/China, Oct'17: (i) MFtD construction with DM curves, sovereign curve, and EM curves; (ii) EM curves with relevant bonds shown.}
\label{fig:10}
\end{figure}

\begin{figure}[!h]
\centering
\begin{tabular}{ll}
(i) & \includegraphics[scale=0.9]{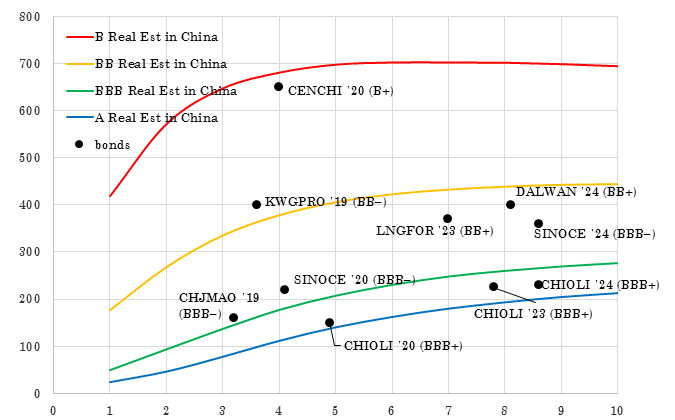} \\
(ii) & \includegraphics[scale=0.9]{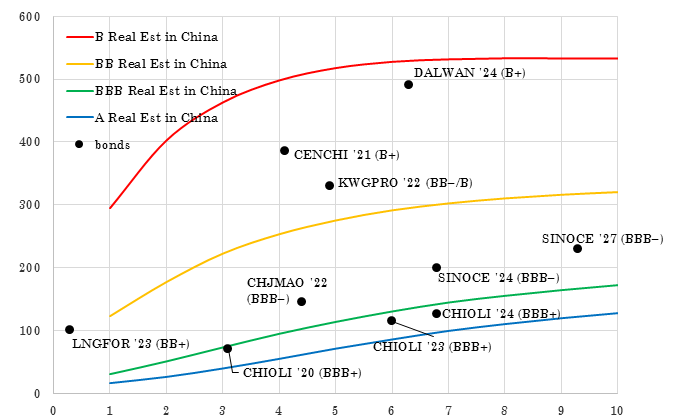} \\
\end{tabular}
\caption{Real Estate/China, (i) Jan'16 and (ii) Oct'17: modified EM curves ($L_a^*=0.85$, see text) with relevant bonds shown.
}
\label{fig:11}
\end{figure}

\end{document}